\documentclass[a4paper]{jpconf}
\usepackage{graphicx}
\usepackage[rightcaption]{sidecap}
\usepackage{cite}
\begin{document}
\title{What perspectives for the synthesis of heavier superheavy nuclei? Results and comparison with models}

\author{G Mandaglio$^{1,2,3}$,  A K Nasirov$^{4,5}$,  F Curciarello$^{1,2}$, V~De~Leo $^{1,2}$,   M. Romaniuk$^{1,2}$G Fazio$^{1,2}$,G Giardina$^{1,2}$} 

\address{$^1$ Dipartimento di Fisica e di Scienze della Terra, Universit\`a di Messina, I-98166 Messina, Italy\\ 
  $^2$  Istituto Nazionale di Fisica Nucleare, Sezione  di Catania, I-95123 Catania,  Italy\\
 $^3$ Centro Siciliano di Fisica Nucleare e Struttura della Materia, I-95123 Catania,  Italy\\ 
   $^4$Joint Institute for Nuclear Research, 141980, Dubna, Russia \\
   $^5$ Institute of Nuclear Physics, 100214,  Tashkent, Uzbekistan}

\ead{giardina@nucleo.unime.it}

\begin{abstract}
The possibility to synthesize heavier superheavy elements in massive nuclei reactions is strongly limited by  the hindrance to complete fusion of reacting nuclei: due to the onset of the quasifission process in the entrance channel, which competes with complete fusion, and by strong increase of the fission yield along the de-excitation cascade of the compound nucleus in comparison to the evaporation residue formation.
We present a wide and detailed procedure allowing us to describe the experimental results (evaporation residue nuclei and fissionlike products) in the mass asymmetric and symmetric reactions. Very reliable estimations and perspectives for the synthesis of superheavy elements in many massive nuclei reactions up to $Z=120$ and eventually also for $Z>120$ have been obtained.
\end{abstract}

\section{Introduction and status}
\label{intro}
Many Laboratories  are strongly engaged to investigate  massive nuclei reactions with the aim to analyze and understand the characteristics and variety of reaction dynamics, and then to plane new experiments for the synthesis of other heavier superheavy nuclei.
In the last decade many superheavy elements with $Z > 110$ were successfully reached by cold and hot fusion reactions, but results of the investigations in some other cases of symmetric or almost symmetric massive nuclei reactions the investigations were unsuccessful. 
New experiments have been performed to synthesize superheavy elements with $Z=120$ and other massive nuclei using reactions being believed to be able to reach superheavy elements with $Z>120$.

The possibility of synthesis of new elements with Z=120, 122, 124,
126 was explored in some hot-fusion reactions (for example the $^{54}$Cr+$^{248}$Cm,
$^{54}$Cr+$^{249}$Cf, $^{58}$Fe+$^{249}$Cf, and $^{64}$Ni+$^{249}$Cf reactions) in cold-fusion reactions (for
example the $^{132}$Sn+$^{174}$Yb, $^{132}$Sn+$^{176}$Hf, $^{132}$Sn+$^{186}$W and $^{84}$Kr+$^{232}$Th reactions) which could lead
to the formation of nuclei in the Z=120-126 range.  Moreover, 
various studies
were conducted by different authors\cite{Siwek07,Swiatecki04,Smolanczuk01,Zagreb07}
 in mass symmetric and asymmetric reactions
($^{136}$Xe+$^{136}$Xe, $^{149}$La+$^{149}$La, $^{86}$Kr+$^{208}$Pb, $^{58}$Fe+$^{244}$Pu) estimating relevant or promising results for the synthesis of superheavy elements, but in the some conducted experiments no events were found\cite{Gregorich03,Ogan09,Ogan2009}. Since some laboratories are planning to perform experiments in such field of nuclear reactions, the present study can be  an useful support of knowledge before to attempt some difficult tasks.
Therefore, it is needed to investigate the conditions and limits of reactions in respect to form compound nuclei (CN), and to produce evaporation residues of superheavy elements. 
 There are three  reasons causing a hindrance to the evaporation residue formation in the
reactions with massive nuclei: the quasifission,  fusion-fission, and fast fission processes\cite{nasirov09,fazio05,fazio08}. 
The quasifission
process competes with the fusion process during the evolution of the dinuclear system (DNS). This process occurs when the
dinuclear system prefers to break down into fragments instead of to be transformed into fully
equilibrated CN. The number of events going to quasifission increases drastically by
increasing the sum of the Coulomb interaction and rotational energy in the entrance channel.
The next reason decreasing yield of ER is the fission of a heated and rotating CN which is
formed in competition with quasifission. The stability of a massive CN decreases due to the
decrease of the fission barrier by increasing its excitation energy $E^*_{\rm
CN}$ and angular momentum $\ell$. The stability of the transfermium nuclei are connected with the availability of shell correction in their binding energy which are sensitive to $E^*_{\rm CN}$ and values of the angular momentum.  
Moreover, the other reason decreasing yield of ER is the fast fission process which is the inevitable decay of the fast rotating mononucleus into the two fragments without reaching the equilibrium compact shape of a CN. Such a mononucleus is formed from  DNS which survived against quasifission at large values of the orbital angular momentum decreasing the fission barrier up to zero. So, the main channels decreasing the cross section of compound nucleus are quasifission and fast fission. These channels produce binary fragments which can overlap with the ones with the fusion-fission channel and the amount of the mixed detected fragments depends on the mass asymmetry of entrance channel, beam energy, as well as the shell structure of being formed reaction fragments. Therefore, the experimental method to extract the fusion-fission contribution by the analysis of the mass and angular distributions of binary fragments of the full momentum transfer events is not unambiguous. 

The failure of many experimental results is connected not only with the difficulties in the measurement of the evaporation residue cross sections which are lower than 0.5 pb but also in the inadequacy of the probability estimation of the complete fusion \cite{Siwek07,Smolanczuk01,Zagreb07} and then in determination of the evaporation residue cross section. The reported difficulties  are related not only with the theoretical estimation of the complete fusion and evaporation residue cross section but also in the not univocal experimental identification of fusion-fission fragments among the quasifission and fast fission fragments. We will also discuss about the limits of reaching compound nuclei  heavier than $Z=120$ due to the dominant repulsive Coulomb effects and strong centrifugal forces in very massive nuclei reactions.

In order to give realistic estimations of cross sections of the reaction products  by mass symmetric or almost symmetric entrance channel it is need to develop an adequate model allowing one to describe by a likelihood way the complex dynamics of the mechanisms during all stages of reaction. 
In fact, in the last stage of  nuclear reaction, the formed CN may de-excite by fission (producing fusion-fission fragments) or by emission of light particles. The reaction products that survive fission are the evaporation residues (ER)\cite{epja222004,fazio05}.
The registration of ER is clear evidence of the CN formation, but in case of reactions with massive nuclei, generally, the knowledge about ER's only  it is not enough to determine the complete fusion cross section and to understand the dynamics  of the de-excitation cascade of CN if the true fission fragments are not included into consideration.  On the other hand, the correct identification of an evaporation residue nucleus by the observation of its $\alpha$-decay chain does not assure if the target material contains other isotopes of the nucleus under consideration. In fact, for example, in the case of the $^{48}$Ca+$^{249}$Cf reaction, the identification of the $^{294}$Hs nucleus as the evaporation residue of the $^{297}$Hs compound nucleus after the emission of 3 neutrons (see the experiment reported in Ref. \cite{Ogan2006}) cannot assure that the collected events of the obtained $^{294}$Hs nucleus are only due  to the mentioned process regarding the $^{297}$Hs  CN formation because also the $^{250}$Cf nucleus, that is inevitably present in the target due to the finite resolution of the mass separation, contributes by the $^{48}$Ca+$^{250}$Cf reaction (leading to the $^{298}$Hs CN) to the synthesis of the same $^{294}$Hs evaporation residue nucleus after 4 neutrons emission from CN. This effect changes with the beam energy and the $E^*_{\rm CN}$ excitation energy of CN.
In addition, the use of some assumptions in separation of the fissionlike fragments according to the kinds of the mechanism of its origin does not allow for sure correct determination of the fusion-fission contribution in the case of overlapping of the mass fragment distributions of different processes (quasifission, fast fission and fusion-fission). The exigence and importance to have a multiparameter and sensitive model is strongly connected with the requirement to reach reliable results and with the possibility to give reliable estimations of perspectives for  the synthesis of superheavy elements. If the estimations reported in Figs. \ref{fig3} (a) and (b) of Ref. \cite{Zag12} about evaporation residue cross section after 2n, 3n, and 4n emission which are peaked at about the same $E^* = 40$~MeV of the $^{298}$116 excitation energy are reliable results, then immediately arises the question: what process and barriers can describe with appreciable probabilities  the emission of 2 and 3 neutrons that take away about 43 and 48 MeV (or also more) of excitation energy from the  $^{298}$116 and $^{299}$118 compound nuclei, respectively? These results mean that each neutron in the evaporation of 2n is emitted with a kinetic energy of about 15 times the nuclear temperature of the   $^{298}$116  CN, and 10 times the nuclear temperature of $^{299}$118 CN  in the 3n evaporation process. That is a fully unrealistic result.  

Moreover, by observing the fission products of fissile nuclei formed 
 in neutron (or very light particles) - induced reactions  on fissile targets, one can conclude that the low excited  compound nucleus (at about $E_{\rm CN}^* <$ 10 MeV) decays into very asymmetric fission fragments (near to the shell closure), while the actinide nuclei or compound nuclei formed in heavy ion collisions at intermediate or high excitation energy ($E_{\rm CN}^* >$ 15 MeV) undergo fission by symmetric fragments. 
Starting from these general observations some researchers put forward the idea that the complete fusion process of two colliding nuclei may be considered as the inverse process of fission. The   authors in the papers \cite{prc782008,Ogan09}  argued that since the fission of a compound nucleus in heavy ion collisions produces just symmetric fragments, then in the collisions of two symmetric (or almost symmetric) nuclei  complete fusion has to be a very probable process. 
But, unfortunately this is not true.  For  systems of colliding nuclei heavier than $^{110}$Pd+$^{110}$Pd fusion does not occur absolutely, while for reactions as  $^{100}$Mo+$^{100}$Mo, $^{96}$Zr+$^{96}$Zr, $^{96}$Zr+$^{100}$Mo, $^{100}$Mo+$^{110}$Pd or induced by projectiles heavier than Zn, Ge, Kr  a strong hindrance to fusion appears. 

Following  the previous reasoning one can affirm that the hypothetical $^{132}$Sn+$^{120}$Cd reaction should  lead to the $^{252}$Cf CN  since  $^{120}$Cd (with $Z$=48 near the shell closure 50) and $^{132}$Sn (with double shell closure $Z$=50 and $N$=82) are produced with high yields in spontaneous fission of $^{252}$Cf. But calculation for this reaction does not give meaningful fusion probability ($P_{\rm CN}<5\times 10 ^{-7}$). 
The simple reason resides in the peculiarities of the reaction dynamics. In the spontaneous fission of $^{252}$Cf the average value of angular momentum distribution of the fragments is close to zero, but if we want to reach the  $^{252}$Cf compound nucleus,  by the hypothetical $^{132}$Sn+$^{120}$Cd reaction (or by the realistic $^{132}$Sn+$^{116}$Cd reaction leading to the  $^{248}$Cf CN), the average value of angular momentum distribution of DNS in the entrance channel may be about $<\ell>=50 \hbar$ or higher by increasing the beam energy. 
In this case the whole peculiarities of the reaction mechanism in the first stage of reaction  should lead almost completely to quasifission (re-separation of nuclei of DNS), while the fusion probability $P_{\rm CN}$ should be lower than about 10$^{-7}$. Morever, the excited and fast rotating deformed mononucleus, which is formed in complete fusion with low probability, undergoes fast fission before the system can reach the compact shape of compound nucleus.

Also in the cases of the explored $^{22}$Ne+$^{250}$Cf (more mass asymmetric  system), $^{24}$Mg+$^{248}$Cm, $^{28}$Si+$^{244}$Pu, $^{34}$S+$^{238}$U, and $^{40}$Ar+$^{232}$Th (less mass asymmetric  system) reactions, the  $^{272}$Hs compound nuclei formed in the different entrance channels at a defined excitation energy $E_{\rm CN}^*$ have different angular momentum distributions. Therefore, the decay rates of CN formed in these reactions will be different. The mass distribution of fusion-fission fragments are peaked at around the $^{136}$Xe nucleus with different dispersions and average angular momentum distributions in connection with the various entrance channels. If we calculate the formation probability of the $^{272}$Hs  compound nucleus in  the mass symmetric $^{136}$Xe+$^{136}$Xe reaction at the same fixed excitation energy $E^*_{\rm CN}$ as in the considered $^{22}$Ne+$^{250}$Cf reaction (where $P_{\rm CN}\simeq$ 1), we do not meaningfully  reach such a compound nucleus($P_{\rm CN}<$10$^{-10}$).  The angular momentum distribution for the $^{136}$Xe+$^{136}$Xe  collision at the capture stage is completely different and all conditions of reaction dynamics  lead to deep inelastic and quasifission products. 
In this context, for  the $^{136}$Xe+$^{132}$Sn  ($P_{\rm CN}<$10$^{-8}$) and $^{132}$Sn+$^{176}$Yb  ($P_{\rm CN}<$5$\times$10$^{-11}$) reactions, one can observe the same above-described hindrance to  complete fusion.

\section{Model and  formalism}

\label{sec:1}

According to the DNS model\cite{Volkov}, the first stage of reaction is the formation of the DNS after full momentum transfer of the relative motion of colliding nuclei into a rotating and excited system. In the deep inelastic collisions DNS is formed for the relatively short time and the full momentum transfer does not occur. Therefore, the deep inelastic collisions are not capture reactions.

The partial capture cross section at a given energy $E_{\rm c.m.}$ and
orbital angular momentum $\ell$ is determined by the formula:
 \begin{equation}
 \label{parcap}
 \sigma^{\ell}_{cap}(E_{\rm c.m.})=
 \pi{\lambda\hspace*{-0.23cm}-}^2
{\cal P}_{cap}^{\ell}(E_{\rm c.m.}) ,
 \end{equation}
where ${\cal P}^{\ell}_{cap}(E_{\rm c.m.})$  is the capture
probability for the colliding nuclei to be trapped into the well of the
nucleus-nucleus potential after dissipation of  part of the initial
relative kinetic energy and orbital angular momentum. The capture probability 
${\cal P}_{cap}^{\ell}$  is equal to 1 or 0 for a given $E_{\rm c.m.}$ energy and
orbital angular momentum $\ell$. Our calculations showed that, depending 
on the center-of-mass system energy $E_{\rm c.m.}$,
there can be ``window'' in  the orbital angular momentum for
capture with respect to the following conditions \cite{fazio05,NPA759}:
  \[{\cal P}^{\ell}_{cap}(E_{\rm c.m.}) = \left\{
  \begin{array}{ll} 1,  \hspace*{0.2 cm}  \rm{if}
\ \  \ell_{min} \leq \ell \leq \ell_d \ \
\rm{and} \ \ {\it E_{\rm c.m.}>V}_{Coul}
  \\ 0, \hspace*{0.2cm} \rm {if}\ \ \ell<\ell_{min}   \ \
  or \ \  \ell>\ell_d  \  \rm {and} \ \
  {\it E_{\rm c.m.}>V}_{Coul}
  \\ 0,  \hspace*{0.2cm} \rm{for \  all} \ \ell \ \
     \hspace*{0.2cm} \rm{if} \ \
  {\it E_{\rm c.m.}\leq V}_{Coul}\:.
 \end{array}
 \right.
 \]
The boundary values $\ell_{min}$ and $\ell_d$ of the partial waves
leading to capture depend on the dynamics of collision and they
are determined by solving the equations of motion for the relative distance
$R$ and orbital angular momentum $\ell$ \cite{epja192004,Giar00,FazioJPSJ}.
At lower energies,  $\ell_{min}$ decreases  to zero and we do not 
observe the $\ell$ ``window'': $0\le\ell\le\ell_d$.
The range of the $\ell$ ``window'' is defined by the size of the
potential well of the nucleus-nucleus
potential $V(R,Z_1,Z_2)$ and the values of the radial $\gamma_R$ and
tangential $\gamma_t$ friction coefficients, as well as by the moment
of inertia for the relative motion\cite{NPA759,Giar00}.

The quasifission process competes with formation of complete fusion. 
This  process occurs when the dinuclear system
prefers to break down into fragments instead of to be transformed
into fully equilibriated  CN.

The fusion excitation function is  determined by product of the partial
capture cross section $\sigma^{\ell}_{cap}$ and the fusion probability
 $P_{CN}$ of DNS at various $E_{\rm c.m.}$ values:
 \begin{equation}
 \label{totfus}
 \sigma_{fus}(E_{\rm c.m.};\beta_P, \alpha_T) = \sum_{\ell=0}^{\ell_f}(2\ell+1)
  \sigma_{cap}(E_{\rm c.m.},\ell;\beta_P, \alpha_T)
 P_{CN}(E_{\rm c.m.},\ell; \beta_P, \alpha_T). 
 \end{equation}

 Consequently, the quasifission cross section is defined by
 \begin{equation}
 \label{totqfis}
 \sigma_{qfis}(E_{\rm c.m.}; \beta_P, \alpha_T)=
 \sum_{\ell=0}^{\ell_d}(2\ell+1)\sigma_{cap}(E_{\rm c.m.},\ell; \beta_P, \alpha_T)
 (1-P_{CN}(E_{\rm c.m.},\ell; \beta_P, \alpha_T)).
 \end{equation}

For more specific details and descriptions on the model see Refs.\cite{fazio05,NPA759,nasirov09,fazio08,faziolett}. 

The fast fission cross section  is calculated
by summing the  contributions of the partial waves corresponding to the
range $\ell_f\le\ell\le\ell_d$ leading to the formation of the mononucleus where the fission barrier $B_{\rm f}$ is zero in such range of $\ell$ ($B_{\rm f}=0$ for $\ell \ge \ell_{\rm f}$) and the nuclear system promptly decays into two fragments:
 \begin{equation}
 \label{fasfiss}
 \sigma_{fast fis}(E_{\rm c.m.};\beta_P,\alpha_T)=\sum_{\ell_f}^{\ell_d}(2\ell+1) \sigma_{cap}(E_{\rm c.m.},\ell; \beta_P, \alpha_T)
 P_{CN}(E_{\rm c.m.},\ell; \beta_P, \alpha_T).
 \end{equation}

The capture cross section in the framework of the DNS model
is equal to the sum of the quasifission,
fusion-fission, and fast fission cross sections:
 \begin{equation}
  \label{capture}
 \sigma^{\ell}_{ cap}(E_{\rm c.m.};\beta_P, \alpha_T)  =
 \sigma^{\ell}_{ qfiss}(E_{\rm c.m.};\beta_P, \alpha_T)
 +\sigma^{\ell}_{ fus}(E_{\rm c.m.}; \beta_P, \alpha_T)
  + \,\,\sigma^{\ell}_{ fast fis}(E_{\rm c.m.}; \beta_P, \alpha_T).
 \end{equation}
It is clear that the fusion cross section includes the cross sections of evaporation residues and fusion-fission products.
The fission cross section is calculated by the advanced statistical
code\cite{dar92,ASM,Sag98} that takes into account the damping
of the shell correction in the fission barrier as a function of 
nuclear temperature and orbital angular momentum.
\begin{equation}
\label{evapor}
\sigma_{ER(x)}(E^*_x)=\sum_{\ell=0}^{\ell_d}(2\ell+1)
\sigma^{\ell}_{(x-1)}(E^*_x)W_{{\rm sur}(x-1)}(E^*_x,\ell),
\end{equation}

The model is also able to calculate the $P_{\rm CN}$ complete fusion probability, mass- angle- kinetic energy- distributions of quasifission and fusion-fission fragments, anisotropy of angular distribution, and the dependence  of cross sections,  Coulomb barrier, intrinsic fusion barrier and quasifission barrier on the orientation angle of colliding nuclei (see Refs. \cite{faziolett,prc84,jpcs2011}).

\section{Sensitivity of the model}

\begin{figure}[h]
\vspace*{4.3cm}
%fig1
\centering
\resizebox{1.05\columnwidth}{!}{\includegraphics{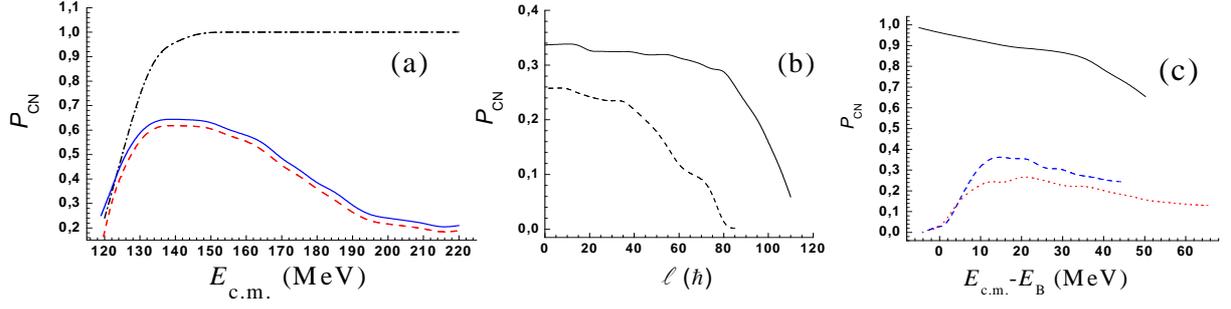}}
\vspace*{-13.1cm} \caption{ The $P_{\rm CN}$ fusion probability calculation: against the $E_{\rm c.m.}$ energy (panel (a)) for the $^{32}$S+$^{184}$W (full line) and $^{32}$S+$^{182}$W (dashed line) reactions, in comparison with the calculation (dash-dotted line) of Ref. \cite{prc782008}; against the angular momentum $\ell$ ($\hbar$) (panel (b)) for the   $^{48}$Ca+$^{154}$Sm reaction  at $E^*_{\rm CN}=49$~MeV (dashed line) and $E^*_{\rm CN}=63$~MeV (full line) excitation energies of the compound nucleus; against the ($E_{\rm c.m.}-E_{\rm B}$) collision energy relative to the interaction barriers (panel (c)) for the $^{16}$O+$^{186}$W (full line) and $^{48}$Ca+$^{154}$Sm (dotted line) reactions leading the  $^{202}$Pb CN, and $^{48}$Ca+$^{144}$Sm (dashed line, very neutron deficient compound nucleus) leading to the $^{192}$Pb CN.\label{fig1}}\vspace{-0.6cm}
\end{figure}

In order to show the sensitivity of our model, in Fig. \ref{fig1}(a), we present (a) the results of $P_{\rm CN}$ fusion probability vs. $E_{\rm c.m.}$ energy for two close reactions leading to $^{214,216}$Th compound nuclei and comparison with very different results of calculation of Ref. \cite{prc782008}; in Fig. \ref{fig1}(b) the results of calculation of $P_{\rm CN}$ vs. the angular momentum values $\ell (\hbar)$ at two $E^*_{\rm CN}$ excitation energy values of the $^{202}$Pb CN by the $^{48}$Ca+$^{154}$Sm reaction; in Fig. \ref{fig1} (c) the results of calculation of $P_{\rm CN}$ vs. ($E_{\rm c.m.}-E_{\rm B}$) collision energy relative to the interaction barrier for two reactions (first and third) leading to $^{202}$Pb and the second reaction leading to $^{192}$Pb (very neutron deficient nucleus).

\begin{figure}[h]
\vspace*{2.3cm}
%fig1
\centering
\resizebox{1\columnwidth}{!}{\includegraphics{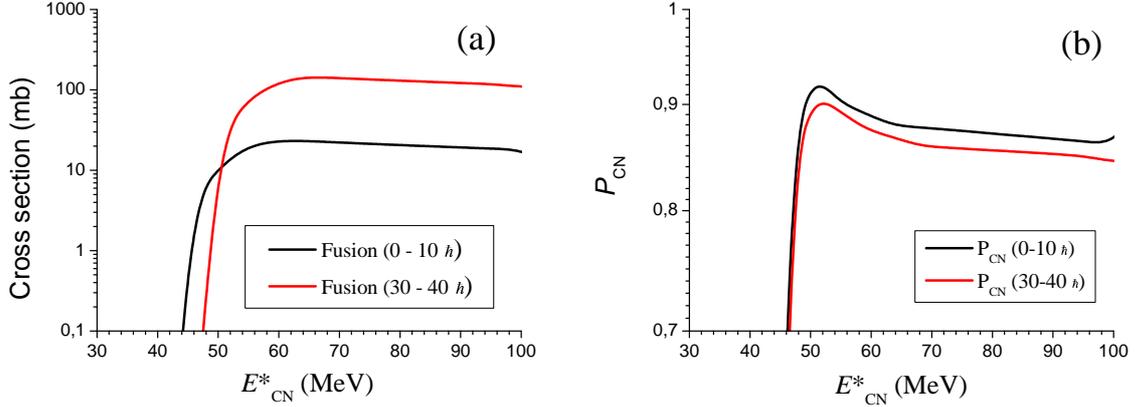}}
\vspace*{-8.4cm} \caption{ Fusion cross section  (panel (a), vs. the $E_{\rm c.m.}$ energy) and  $P_{\rm CN}$ values (panel (b), vs. the $E^*_{\rm CN}$ excitation energy of CN) for the $^{24}$Mg+$^{248}$Cm reaction and an angular momentum range of 10~$\hbar$ taken around 5~$\hbar$ (black line) and 35~$\hbar$ (red line). 
\label{fig2}}\vspace{1.6cm}
\end{figure}
Moreover, Fig. \ref{fig2} (a) shows the comparison  of the fusion cross section determinations vs. the $E_{\rm c.m.}$ energy for one fixed interval of $10\hbar$ of the angular momentum   selected between $(0-10)\hbar$ and $(30-40)\hbar$, for the $^{24}$Mg+$^{248}$Cm reaction. An analogous description is also obtained for the results of the capture cross section. In Fig. \ref{fig2} (b) we report values of the $P_{\rm CN}$ fusion probability determined for the two considered angular momentum intervals, as a function of the  $E^*_{\rm CN}$ excitation energy of the $^{272}$Hs CN. As one can see, $P_{\rm CN}$ appreciably changes  (remaining about 0.85) at  $E^*_{\rm CN}>50$~MeV for this investigated reaction.
\begin{figure}[h]
\vspace*{-0.6cm}
%fig1
\centering
\resizebox{1.1\columnwidth}{!}{\includegraphics{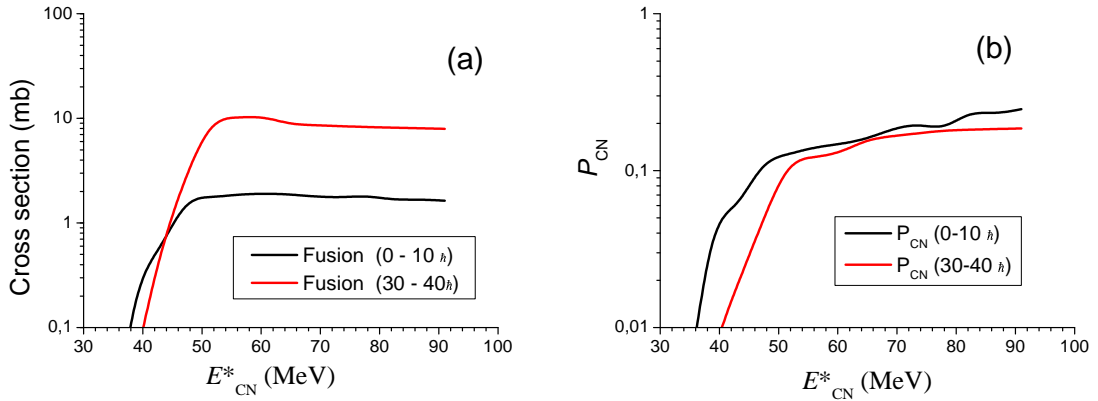}}
\vspace*{-8.7cm} 
\caption{As Fig. \ref{fig2}, but for the $^{34}$S+$^{238}$U reaction. \label{fig3}}\vspace{-0.6cm}
\end{figure}
Figure \ref{fig3} (a) shows analogous results for the fusion cross section as shown in Fig. \ref{fig2} (a) but here presented the ones calculated for the  $^{34}$S+$^{238}$U reaction. In Fig. \ref{fig3} (b) we report the results of calculation of $P_{\rm CN}$ as presented in Fig. \ref{fig2} (b) but here it shown results obtained for the $^{34}$S+$^{238}$U reaction. As one can see, also for this reaction $P_{\rm CN}$ appreciable changes  at $E^*_{\rm CN}>55$~MeV (remaining yet at about 0.15), while at lower excitation energy the  $P_{\rm CN}$ values  strongly change at least by about factor 5. The results of Figs. \ref{fig1}, \ref{fig2}, and \ref{fig3} demonstrate  the importance of  including  the dependence on the angular momentum distributions in order to obtain  reliable calculation of cross sections or other functions  characterizing process. Therefore, the methods of calculation that do not take into account  the dependences of the $P_{CN}$ fusion probability on the angular momentum $\ell (\hbar)$, $E_{c.m.}$ energy, and orientation angles of the symmetry axes of deformed  reacting nuclei cannot reach complete understanding of the peculiarities of the fusion mechanism.

\section{Study of the $^{50}$Ti+$^{249}$Cf and $^{54}$Cr+$^{248}$Cm reactions}
\begin{figure}
% Fig.5
\vspace*{-1.5cm}
\begin{center}
\resizebox{1.00\textwidth}{!}{\includegraphics{{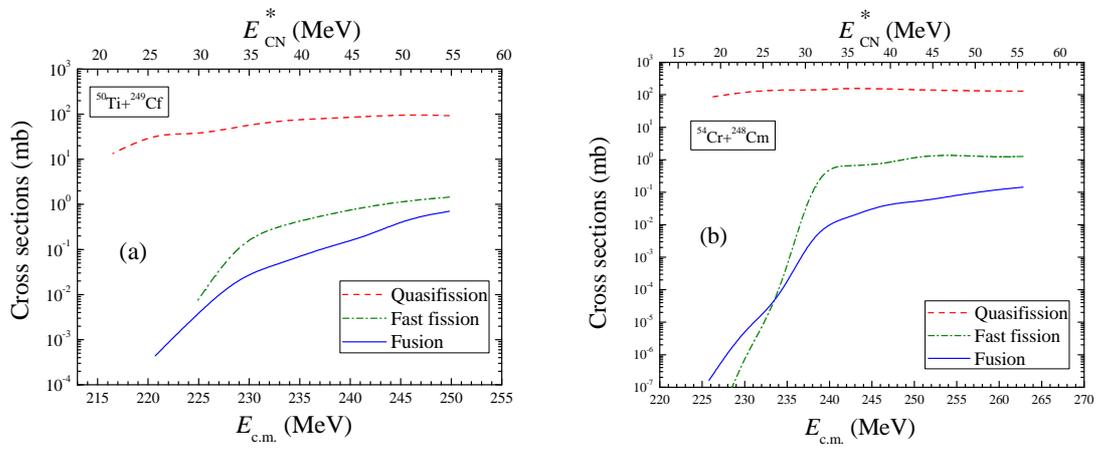}}}% Here is how to import EPS art
\vspace*{-7.52 cm} \caption{\label{CapFusTiCf} (a) Quasifission (dashed line),
fast fission (dot-dashed line), and complete fusion (solid line) excitation functions
calculated by the DNS model \cite{faziolett,NPA759,fazio05} for the
$^{50}$Ti+$^{252}$Cf reaction  which could lead to the $^{299}120$ compound nucleus.(b) The same as (a), but for the $^{54}$Cr+$^{248}$Cm reaction  which could lead to the $^{302}120$ compound nucleus.}
\end{center}
\end{figure}

 In Figs. \ref{CapFusTiCf} (a) and (b), we present our theoretical results for
 quasifission, fast fission and complete fusion cross sections of the $^{50}$Ti+$^{249}$Cf
  and $^{54}$Cr+$^{248}$Cm  reactions.
  The comparison of these figures shows that at low energies capture cross sections
 of the $^{54}$Cr+$^{248}$Cm reaction is larger  than ones of the $^{50}$Ti+$^{249}$Cf reaction,
  while these cross sections become comparable at more large energies corresponding to the 3n- and 4n-channel.
  As one can see that the fusion cross section is
  sufficiently larger for the $^{50}$Ti+$^{249}$Cf reaction in comparison with  the one of
  the $^{54}$Cr+$^{248}$Cm reaction.
 The advance of the charge asymmetric system  appears at the second stage (fusion) of the
 reaction mechanism leading to formation of the evaporation residues.
 It is well known that the hindrance to complete fusion decreases by increasing the DNS
 charge asymmetry. At the same time the DNS quasifission barrier, $B_{\rm qf}$,
 increases because the Coulomb repulsion forces decrease by  decreasing  the 
 $Z_1\cdot Z_2$ product.
  Therefore, in spite of the fact that  the
  $^{50}$Ti+$^{249}$Cf system which has less neutrons in comparison with $^{54}$Cr+$^{248}$Cm,
  the probability of compound nucleus formation is higher for the former reaction than
  for the latter reaction. The more strong hindrance in  the case of the
  $^{54}$Cr+$^{248}$Cm reaction is connected with the larger intrinsic fusion barrier
  $B^*_{\rm fus}$ and smaller quasifission barrier $B_{\rm qf}$ for this reaction in
  comparison with $^{50}$Ti+$^{249}$Cf.
  
\begin{figure}
% Fig.7
\vspace*{1.0cm}
\begin{center}
\resizebox{1.00\textwidth}{!}{\includegraphics{{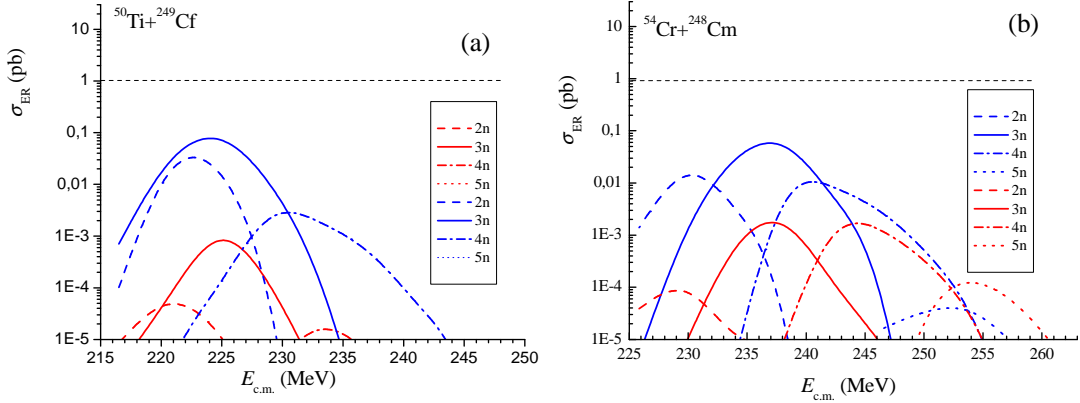}}}% Here is how to import EPS art
\vspace*{-7.50 cm} \caption{\label{ERTiCf} 
Comparison of the evaporation residue excitation functions
for the $^{50}$Ti+$^{252}$Cf reaction calculated by using mass tables of Nix-Moeller
\cite{MolNix} (thick blue lines) and Sobiczewski's group \cite{Muntian03,Kowal} (thin red lines)
for the 2n (dashed lines), 3n (solid lines), 4n (dot-dashed lines), and 5n
(dotted lines) channels calculated by the advanced statistical model
\cite{dar92,ASM,Sag98}. (b) The same as (a) but for the $^{54}$Cr+$^{248}$Cm reaction.
}
\end{center}\vspace*{-1.0cm}
\end{figure}

 The theoretical excitation functions of evaporation residues which can be formed
 in different neutron-emission channels for these two systems are presented in Figs.
 \ref{ERTiCf} (a) and (b). In each of the figures the evaporation  residue cross  sections
  for the neutron-emission channels obtained by using  binding energies and fission barriers
  calculated in the microscopic-macroscopic  models of Nix-M\"oller \cite{MolNix} and
 Sobiczewski's group \cite{Muntian03,Kowal} are compared. The difference between binding energies
 obtained by these two groups is in the range  of  2-3 MeV for the isotopes of superheavy
 nuclei  with $Z > 114$. This difference causes a difference between values
 of the branching  ratios $\Gamma_n/\Gamma_f$  which is used in calculations
 of the survival probability of heated and rotating nuclei.  The use of 
 the mass table  of the Sobiczewski's group  leads to  two main consequences:
 the excitation energy of the compound nucleus will be lower because the negative
 value of $Q_{\rm gg}$ is larger, and the value of the fission barrier $B_f$
 is lower increasing  the fission probability. The evaporation residues
 cross sections obtained by the use of mass table calculated by the
 Nix-M\"oller microscopic-macroscopic  model  are  one order of magnitude larger
 in comparison with  the results obtained   by the use of   the mass table  of  Sobiczewski's group.

\section{Superheavy nuclei and perspective for heavier superheavy elements}
 
 In order to estimate the realistic possibilities to synthesize superheavy elements by massive nuclei reactions, we performed calculations of many reactions forming fissile compound nuclei with $Z\ge$ 100 at the same excitation energy ($E^*_{\rm CN} \simeq$ 37 MeV). In Table 1 we present the set of studied reactions leading to heavy and superheavy elements by various entrance channels with different charge (mass) asymmetry parameters.

\begin{table}[h]
\centering\vspace{-0.2cm}
\begin{tabular}{|c|c|c||c|c|c|}
\hline
Reaction & $Z_{\rm CN}$ & $z$ & Reaction & $Z_{\rm CN}$ & $z$\\
\hline
$^{16}$O+$^{238}$U	& 100	&  84&    $^{86}$Kr+$^{208}$Pb	& 118	& 286\\
\hline
$^{48}$Ca+$^{208}$Pb	& 102	& 172 &    $^{132}$Sn+$^{174}$Yb	& 120	& 328\\
\hline
$^{50}$Ti+$^{208}$Pb	& 104	& 188 &   $^{64}$Ni+$^{238}$U	& 120& 253 \\
\hline
$^{136}$Xe+$^{136}$Xe		& 108	& 284 & $^{58}$Fe + $^{244}$Pu	& 120	& 242\\
\hline
$^{58}$Fe+$^{208}$Pb	& 108	& 218& $^{54}$Cr + $^{248}$Cm	& 120	& 229\\
 \hline
$^{48}$Ca+$^{226}$Ra	& 108	& 181 & $^{132}$Sn+$^{176}$Hf& 122 & 337\\
\hline
$^{26}$Mg+$^{248}$Cm& 	108& 125 & $^{54}$Cr +$^{249}$Cf	&122	&234\\
\hline
$^{48}$Ca+$^{243}$Am	& 115	& 193 & $^{132}$Sn+$^{186}$W	&124	&343\\
\hline
$^{48}$Ca+$^{248}$Cm	& 116	& 194& $^{58}$Fe+$^{249}$Cf	&124	&251\\
\hline
$^{48}$Ca+$^{248}$Bk	& 117	& 196& $^{84}$Kr+$^{232}$Th	&126	&307\\
\hline
$^{48}$Ca+$^{249}$Cf	& 118	& 198 &  $^{64}$Ni+$^{249}$Cf	&126	&267\\
\hline
\end{tabular}
\caption{The listed reactions are reported as a function of the charge $Z_{\rm CN}$ of compound nucleus (if it can be reached), and the parameter $z=\frac{Z_1\times Z_2}{A_1^{1/3}+A_2^{1/3}}$ representing the Coulomb barrier of reacting nuclei in the entrance channel.}
\label{tab1}%\vspace{2cm}
\end{table}

It is interesting to observe and analyze the overall trend of the fusion probability $P_{\rm CN}$ and the evaporation residue yields for various reactions as a function of the charge $Z$ of CN and of the parameter $z=\frac{Z_1\times Z_2}{A_1^{1/3}+A_2^{1/3}}$ (related to the Coulomb barrier in the entrance channel)  in order to draw some useful indications on the possible reactions leading to heavy nuclei with $Z\ge$ 100 and particularly on reactions leading to superheavy elements with $Z\ge$ 120.

\begin{figure}[h]
\vspace*{-1.8cm}
%fig1
\centering%\vspace{3.5cm}
\resizebox{1\columnwidth}{!}{\includegraphics{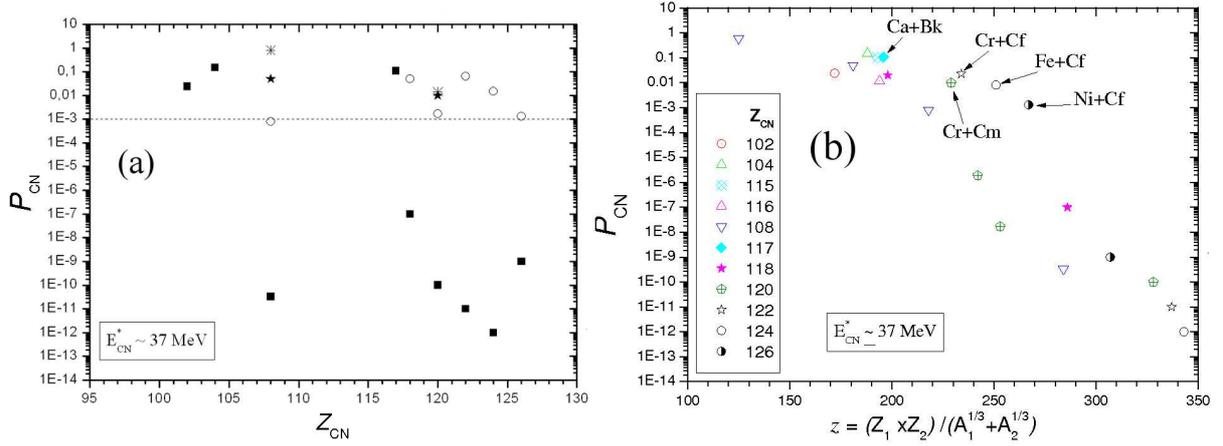}}
\vspace*{-0.62cm} \caption{ (a) Fusion probability $P_{\rm CN}$ versus charge $Z_{\rm CN}$, for the reactions listed in Table 1, calculated  at the same excitation energy $E^*_{\rm CN}\simeq$ 37 MeV. (b) Fusion probability  $P_{\rm CN}$ versus the parameter $z$ (representing the Coulomb barrier of reacting nuclei in the entrance channel) for many reactions with charge of compound nucleus $Z_{\rm CN}$ included in the $Z_{\rm CN}$ = 102-126 range.}\vspace{-0.5cm}
\label{pcnbn}
\end{figure}

Figure \ref{pcnbn}(a) shows the fusion probability $P_{CN}$ for the reactions listed in Table \ref{tab1} as a function of the charge $Z$ of CN, at excitation energy $E^*_{\rm CN}\simeq$ 37 MeV. As one can see in this figure,  $P_{\rm CN}$ slowly decreases with $Z$ but strongly decreases for more symmetric reactions in the entrance channel leading to the same $Z_{\rm CN}$. The trend of $P_{\rm CN}$ for the same investigated reactions  appears more clear if we report the calculated $P_{\rm CN}$ as a function of the parameter $z=\frac{Z_1\times Z_2}{A_1^{1/3}+A_2^{1/3}}$ representing the effect of the Coulomb barrier of interacting nuclei in the entrance channel (see Fig. \ref{pcnbn} (b)). 

The different symbols and  values of $P_{\rm CN}$ reported at the same $Z_{\rm CN}$ (108, 118, 120, 122, 122, 124, 126) represent different fusion probabilities for various entrance channels leading to the same $Z_{\rm CN}$. The $P_{\rm CN}$ values decrease for less asymmetric reactions.
As Fig. \ref{pcnbn}(b) shows the trend of $P_{\rm CN}$ at $E^*_{\rm CN}\simeq$ 37 MeV strongly decreases by  increasing  the $z$ parameter and by  decreasing  the charge (mass) asymmetry parameter of reactions in the entrance channel. The hindrance to fusion  increases for more symmetric reactions and for higher Coulomb barriers of reactions in entrance channel. 
\begin{figure}[h]
\vspace*{-2.9cm}
%fig1
\centering\vspace{2,7cm}
\resizebox{1\columnwidth}{!}{\includegraphics{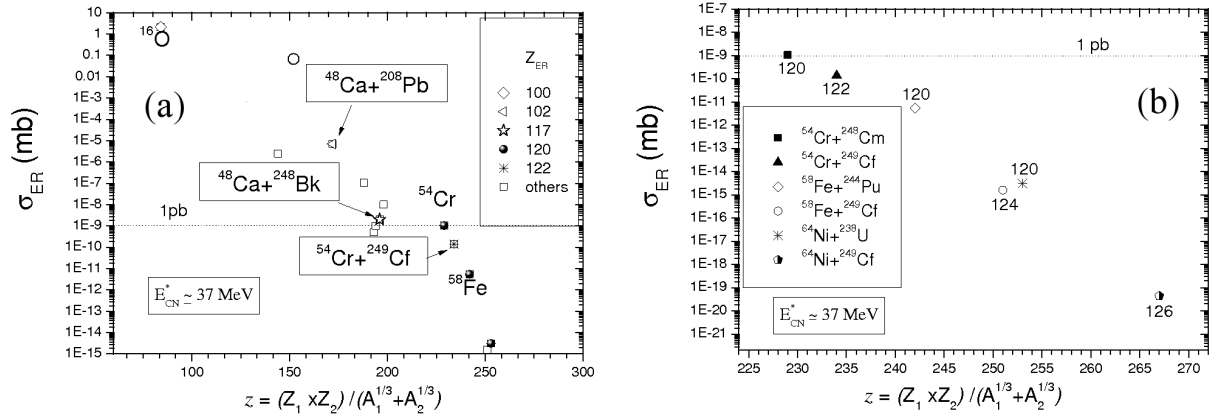}}
\vspace*{-0.72cm} \caption{(a) Evaporation residue cross section $\sigma_{\rm ER}$ (after neutron emission only versus the parameter $z$ representing the Coulomb barrier in the entrance channel, for reaction with $Z_{\rm CN}$= 100-120. (b) As panel (a), but for reactions leading to $Z_{\rm CN}$= 120-126.
\label{ercolor}}\vspace{-0.4cm}
\end{figure}

Figure \ref{ercolor} (a) shows the evaporation residue cross sections, after neutron emission only from CN, obtained for the investigated reactions as a function of the parameter $z$, at  $E^*_{\rm CN}\simeq$ 37 MeV. In the figure the horizontal dotted line marks the value of 1 pb for the ER cross section. One can see that for reactions with parameter $z$ lower than the value of about 200 it is possible to observe evaporation residues after neutron emission only from the de-excitation cascade of the compound nucleus. For reactions with values of parameter $z$ included in the about 200-235 range the observation of residues is at limit (or it appears to be a very problematic task) of the current experimental possibilities. For reactions with $z$ higher than 235 it is impossible to observe ER of CN after neutron emission only. 

\begin{table}[h]
\centering \vspace{0.5cm}
\begin{tabular}{|c|c|c|c|c|}
\hline
Reaction & $Z_{\rm CN}$ & $z$  & $\sigma_{\rm ER}$ (mb) &  $P_{res/cap}$\\
\hline
$^{54}$Cr + $^{248}$Cm		& 120& 	229 &1.05 $\times$ 10$^{-9}$ &  0.3 $\times$ 10$^{-10}$ \\
\hline
$^{58}$Fe + $^{244}$Pu	& 120	& 242& 5.4 $\times$ 10$^{-12}$ & 0.17 $\times$ 10$^{-13}$ \\
\hline
$^{64}$Ni+$^{238}$U & 120	& 253 &  3.1 $\times$ 10$^{-15}$ & 0.14 $\times$ 10$^{-15}$\\
\hline
$^{54}$Cr +$^{249}$Cf	&122	&234&  1.4 $\times$ 10$^{-10}$ & 0.13 $\times$ 10$^{-11}$\\
\hline
$^{58}$Fe+$^{249}$Cf	&124	&251& 1.61 $\times$ 10$^{-15}$ & 0.18 $\times$ 10$^{-16}$ \\
\hline
$^{64}$Ni+$^{249}$Cf	&126	&267& 4.4 $\times$ 10$^{-20}$ & 6.5 $\times$ 10$^{-22}$ \\
\hline
\end{tabular}
\caption{Reactions leading to compound nuclei with $Z_{\rm CN}$= 120-126, as a function of the parameter $z$ representing the Coulomb barrier in the entrance channel.  $\sigma_{\rm ER}$ is the ER cross section after the neutron emission only from the de-excitation cascade of CN; $P_{res/cap}$ is the ratio between the yields of evaporation residue $\sigma_{\rm ER}$  and the capture $\sigma_{\rm cap}$.}
\label{tab:}\vspace{-0.4cm}
\end{table}
We report in Table \ref{tab:} the results obtained for the investigated reactions leading to CN with $Z=$120, 122, 124 and 126, at excitation energy of compound nuclei of about 37 MeV. 
Figure \ref{ercolor} (b) shows the results of ER as a function of the  parameter  $z=\frac{Z_1\times Z_2}{A_1^{1/3}+A_2^{1/3}}$, at $E^*_{\rm CN}\simeq$ 37. In the figure is reported by dotted line the value of $\sigma_{\rm ER}$ of 1 pb. 

As one can see we estimate that only for the superheavy element with $Z=$ 120 it is possible to observe evaporation residues by reactions with $z$ parameter lower than 230. 
The observation of the superheavy element   $Z=$ 122 by reaction with  $z$ of about 234
appears to be a very doubtful venture. 
The observation of superheavy elements   with $Z=$ 124 and 126  by reactions with $z$ of about 251 and 267, respectively,  is impossible by the current experimental conditions and detecting system of evaporation residues. 

\section{Conclusions}

We studied the distribution of the fusion probability $P_{\rm CN}$  versus the charge  $Z$ of the compound nucleus for a wide set of reactions by dynamical calculations.  The formation probability of a dinuclear system in the entrance channel  and its subsequent evolution into compound  nucleus  in competition with quasifission is analyzed by  the systematics of $P_{\rm CN}$ versus the parameter z representing the Coulomb barrier in the entrance channel.  The evaporation residue cross sections $\sigma_{\rm ER}$ versus the $z$ parameter calculated for the different reactions are presented. The  ER  cross sections are obtained for the de-excitation cascade of CN after neutron emission only. From the study of such systematics in many reactions forming various compound nuclei at the  same excitation energy $E_{\rm CN}$  of about 37 MeV it is possible to understand the role of the entrance channel mass symmetry in the complete fusion reactions and in formation of  the evaporation residue. 

The trend of $P_{\rm CN}$ is a slow decrease of its values with the increase of the charge  
 $Z$ of compound nucleus and by the decrease  of the mass asymmetry parameter of 
entrance channel of the reactions which lead to form the same compound nucleus, as well as a fast decrease of $P_{\rm CN}$ values and ER yields versus the parameter $z=\frac{Z_1\times Z_2}{A_1^{1/3}+A_2^{1/3}}$.

At conclusion of the present investigation,  the use of the neutron rich radioactive beam $^{132}$Sn  for the formation of superheavy nuclei is not of promising possibilities. 

Regarding the results of the investigated reactions  leading to the formation of compound nuclei with $Z=$ 120, 122, 124 and 126, we affirm that it is possible to reach and observe the ER of the 120 superheavy element by a reaction with $z$ parameter of about 230, while it is a very doubtful venture to synthesize the 122 superheavy element by reactions with $z$ parameter of about 234 or higher by the current experimental resources and methods of observing evaporation residues.

 The possibility to observe evaporation residue of superheavy elements   appears out in reactions with $z$ parameter in the entrance channel higher than 240. Therefore, it is impossible to form the 124 and 126 superheavy nuclei by the studied reactions above mentioned.

The quasifission is the main cause of hindrance of complete fusion and the yield of such a process strongly increases for reactions with higher $z$ parameters and also with the increase of the $E^*_{\rm c.m.}$ energy. The fast fission and fusion-fission are the subsequent hindrances to lead to evaporation residues at forming of complete fusion and reaching of compound nucleus CN.  In this context, the mass symmetric or nearly symmetric reactions in the entrance channels do not give a realistic possibility to synthesize superheavy elements, and the use of the $^{132}$Sn beam is of scarce usefulness for this kind of reactions. 

Consequently, it is an unrealizable  dream to think of performing the $^{132}$Sn+$^{208}$Pb (with $z$ = 373) and  $^{132}$Sn+$^{249}$Cf (with $z$ = 431) reactions in order to reach the $^{340}$132 and $^{381}$148 superheavy elements, respectively, and by mass symmetric reactions like $^{136}$Xe+$^{136}$Xe (with $z$ = 184)  and \\$^{139,149}$La+$^{139,149}$La (with $z$ = 317 and 306, respectively)  to synthesize heavy and superheavy elements to cause of the absolute dominant contribution of the quasifission process after capture, and the fast fission process presents at stage of the little probable formation of complete fusion.   

Moreover, we also presented the results of the $^{50}$Ti+$^{252}$Cf and $^{54}$Cr+$^{248}$Cm reaction leading to the $^{299}$120 and $^{302}$120 compound nuclei, respectively. We discussed about the evaporation residue cross section determinations when the masses and barriers of  Nix-Moeller \cite{MolNix} or Sobiczewski's group \cite{Muntian03,Kowal} are used. 
In addition, we compared and discussed results obtained by other models and explained why  the complete fusion process of the two colliding massive nuclei (for example, in the $^{132}$Sn+$^{120}$Cd reaction) cannot be considered as the inverse process of fission (for example, in the $^{252}$Cf spontaneous fission). We discussed the huge role of the Coulomb forces on the hindrance to complete fusion in the $^{136}$Xe+$^{136}$Xe reaction in comparison with the fully different case of the $^{24}$Mg+$^{248}$Cm reaction leading to high rate of complete fusion of reactants in the entrance channel.

\end{document}